\documentclass[3p,times]{elsarticle}

\usepackage{ecrc}

\volume{00}

\firstpage{1}

\journalname{Nuclear Physics A}

\runauth{X.-N. Wang}

\jid{nupha}

\usepackage{amssymb}

\usepackage[figuresright]{rotating}

\begin{document}

\begin{frontmatter}



\dochead{}

\title{Dihadron and $\gamma$-hadron correlations from jet-induced medium excitation in high-energy heavy-ion collisions}


\author[IOPP]{Han-Lin Li} 
\author[IOPP]{Fu-Ming Liu}
\author[SINAP]{Guo-Liang Ma}
\author[IOPP,LBNL]{Xin-Nian Wang}
\author[IOPP]{Yan Zhu}
\address[IOPP]{Institute of Particle Physics, Hua-Zhong Normal University, Wuhan 430079, China}
\address[SINAP]{Shanghai Institute of Applied Physics, Chinese Academy of Sciences, P.O. Box 800-204, Shanghai 201800, China}
\address[LBNL]{Nuclear Science Division MS 70R0319, Lawrence Berkeley National Laboratory, Berkeley, California 94720}

\begin{abstract}
Jet propagation is shown to produce Mach-cone-like medium excitation inside
a quark-gluon plasma. However, only deflection of such medium excitation and jet shower partons by radial flow leads 
to double-peaked dihadron correlation in high-energy heavy-ion collisions. Dihadron correlations from harmonic flow,
hot spots and dijets are studied separately within the AMPT Monte Carlo model and all lead to double-peaked dihadron
azimuthal correlation. The $\gamma$-hadron correlation has similar double-peak feature but is free of the contributions from
harmonic flow and hot spots. Dihadron and $\gamma$-hadron correlations are compared to shed light
on jet-induced medium excitation and hot spots in an expanding medium.

\end{abstract}


\end{frontmatter}



The quark-gluon plasma produced in heavy-ion collisions at RHIC is very opaque to energetic partons as shown by the observed
strong jet quenching \cite{Adler:2003qi,Adler:2002tq} due to multiple scattering and induced parton energy loss \cite{Wang:1991xy}.
The energy and momentum lost by a propagating parton will be transferred to the medium via
radiated gluons and recoiled partons and lead to collective medium excitation in the form of Mach 
cones \cite{CasalderreySolana:2004qm,Stoecker:2004qu}. Such collective excitation by a propagating jet 
is expected to be responsible for the observed azimuthal dihadron correlation \cite{Adams:2005ph,Adler:2005ee} that has
a double-peak on the back-side of a triggered high-$p_{T}$ hadron. However, hadron spectra from the 
freeze-out of the Mach cone in both hydrodynamics with realistic energy-momentum deposition by 
jets \cite{CasalderreySolana:2004qm} and string calculations in the 
hydrodynamic regime \cite{Gubser:2009sn} fail to reproduce the observed conic azimuthal correlations. 
In this talk, we will report a recent study \cite{Li:2010ts} of medium excitation by a propagating jet shower 
using both a linear Boltzmann transport and AMPT model \cite{Zhang:1999bd}. We will illustrate
that while a Mach-cone-like excitation by a propagating jet in a uniform medium cannot give rise to
a conic distribution of the final partons, deflection of the jet shower and the
Mach-cone-like excitation in an expanding medium will result in a double-peaked azimuthal distribution. 

Recent studies also found that hydrodynamic expansion of hot spots or local fluctuation in the initial 
parton density under the influence of radial flow \cite{Takahashi:2009na} and the triangular flow
of dense matter with fluctuating initial geometry \cite{Alver:2010gr} all contribute to a double-peaked
back-to-back dihadron correlation.  With these different mechanisms
contributing to the dihadron correlation, it is important to explore ways to separate different contributions
and study the characteristics of the dihadron correlation from each of them.  We will also report
a recent study \cite{Ma:2010dv} on dihadron correlation
as a result of harmonic flow or high order azimuthal anisotropy of hadron spectra, 
expanding hot spots, jets and jet-induced medium excitation. The dihadron correlation after subtraction of contributions
from harmonic flow should come from medium modified jets,  jet-induced medium excitation and expanding hot spots under
strong radial flow in high-energy heavy-ion collisions.   By successively
randomizing the azimuthal angle of transverse momenta and transverse coordinates of initial jet shower
partons, we can isolate the effects of medium modified dijets, jet-induced medium excitation and
expanding hot spots. Because of the azimuthal uniform emission of direct photons, $\gamma$-hadron correlation should be free
of contributions from harmonic flow and hot spots and therefore is caused only by jet-induced medium
excitations. We therefore propose to use comparative study of $\gamma$-hadron and dihadron azimuthal correlations
to disentangle contributions from expanding hot spots  and shed light on the dynamics of jet-induced Mach-cone-like 
excitation in high-energy  heavy-ion collisions


One can study jet shower propagation and medium excitation through a linearized Monte
Carlo simulation of the Boltzmann transport equation,
\begin{eqnarray}
 p _1  \cdot \partial f_1 (p_1 ) &=&  - \int {dp_2 } {dp_3 } {dp_4 }   (f_1 f_2  - f_3 f_4 ) 
 \left| {M_{12 \to 34} } \right|^2 \nonumber \\
 &\times& 
   (2\pi )^4 \delta ^4 (p_1  + p_2  - p_3  - p_4 ), 
   \label{eq:boltz}
\end{eqnarray}
including only elastic $1+2\rightarrow 3+4$ processes as given by the matrix elements $M_{12 \to 34}$, 
where $ dp_i \equiv d^3 p_i/[2E_i (2\pi )^3]$, $ f_{i}  =1/(e^{p_{i}\cdot u/T}  \pm 1)$ $(i=2,4)$ are thermal parton 
phase-space densities in a medium with local temperature $T$ and flow velocity $u=(1,\vec v)/\sqrt{1-v^{2}}$, 
$f_{i}=(2\pi)^{3}\delta^{3}(\vec p-\vec p_{i})\delta^{3}(\vec x-\vec x_{i}-\vec v_{i}t)$ $(i=1,3)$ are the jet shower parton
phase-space densities before and after scattering, and we neglect the quantum statistics in the final state 
of the scattering. We will consider quark propagation in a thermal medium and assume small 
angle approximation of the elastic scattering amplitude  $\left| {M_{12 \to 34} } \right|^{2} = C g^{4}(s^2  + u^2
)/(t + \mu^{2 })^2 $ with a screened gluon propagator, where $s$, $t$ and $u$ are Mandelstam variables, 
$C$=1 (9/4) is the color factor for quark-gluon (gluon-gluon) scattering and $\mu$ is the screening mass
which we consider here as a constant cut-off of small angle scattering. The corresponding elastic cross section 
is $d\sigma/dt=\left| {M_{12 \to 34} } \right|^{2}/16\pi s^{2}$.

Our numerical simulations show \cite{Li:2010ts} that recoil thermal partons from jet-medium interaction in a uniform medium
do cause  Mach-cone-like excitation. The low energy recoil partons around the neck of the Mach-cone-like 
excitation have a double-hump feature in the azimuthal distribution. However, low energy partons from the 
body of the Mach-cone-like excitation become dominate at later times and the final azimuthal distribution has 
only a broad single peak along the direction of the propagating jet due to diffusion of the wake front.

\begin{figure}[!ht]
\centerline{\includegraphics[width=6.5cm]{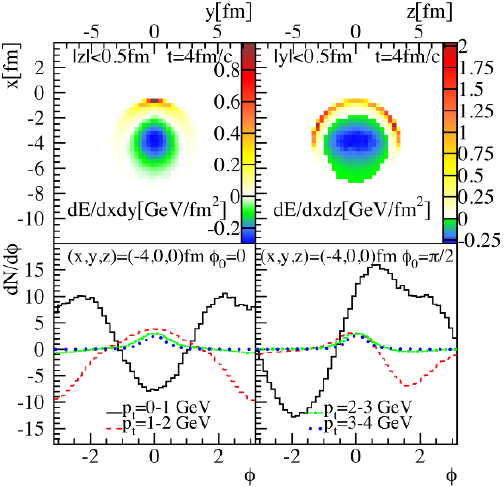}\includegraphics[width=6.5cm]{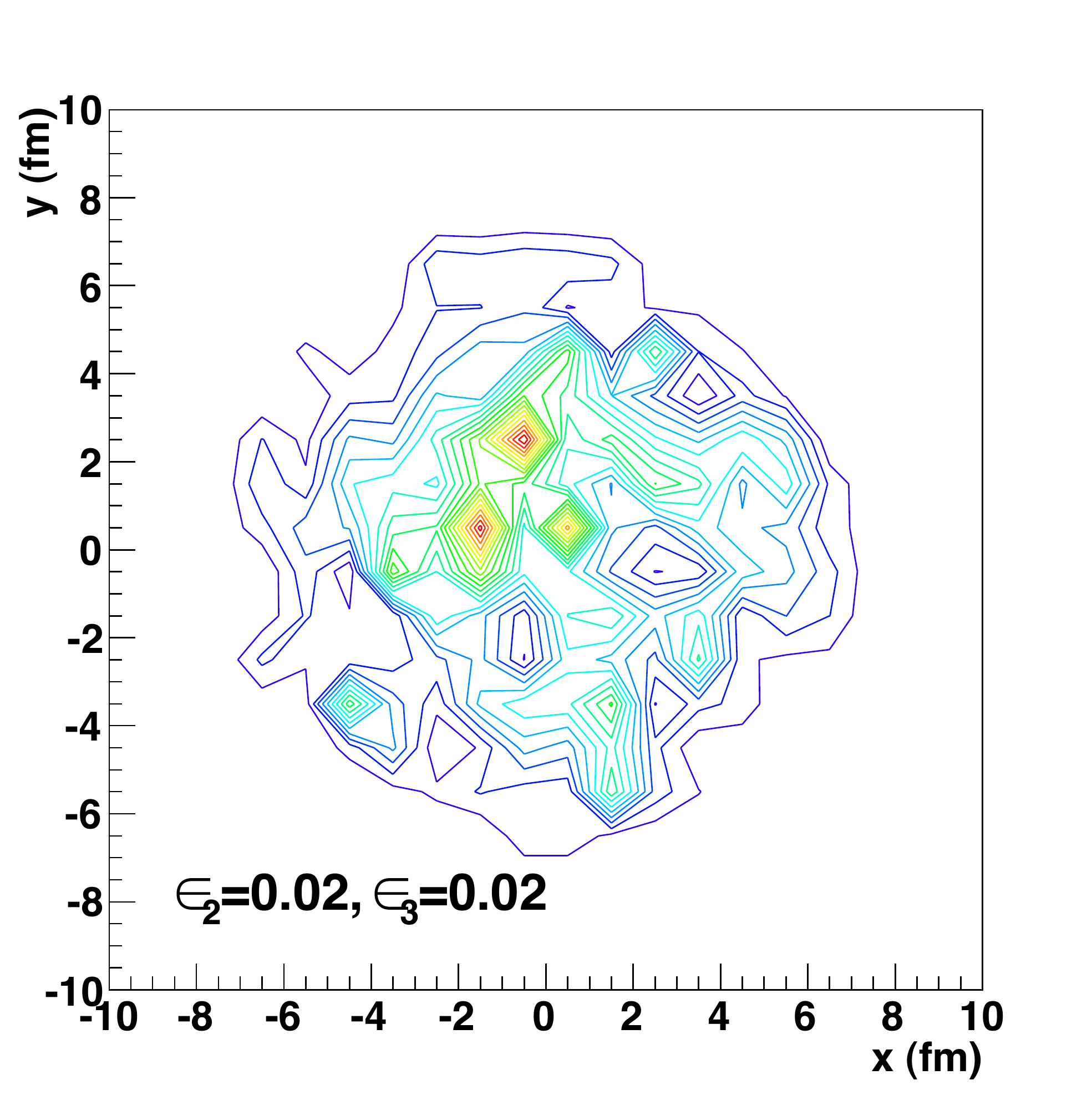}}
\caption{(Color online) (left upper) Contour plot in the transverse ($x$-$y$) and beam ($x$-$z$) plane of energy density
excited by a quark jet shower with $E_{T}=20$ GeV and initial position at $(x,y,z)=(-4,0,0)$ fm and travels toward the center of
the expanding medium. The azimuthal distribution of medium and jet shower partons when the jet shower
travels against (left lower left) and perpendicular (left lower right) to the transverse flow.
(right) Contour plot of initial parton density (in arbitrary unit)  $dN/dxdy$ in transverse
plane in a typical AMPT central $Au+Au$ event ($b=0$) at $\sqrt{s}=200$ GeV, with ellipticity $\epsilon_{2}=0.02$ 
and triangularity $\epsilon_{3}=0.02$ of the transverse parton distribution.}
\label{fig1}
\end{figure}

In an expanding medium as described by  a (3+1)D ideal hydrodynamical calculation \cite{Hirano:2005xf},
both the  shape of the medium excitation and the azimuthal distribution of partons from the jet shower 
and jet-induced medium excitation are distorted by the transverse flow and the non-uniformity of the dense medium
as shown in the left panel of Fig.~\ref{fig1}.  For a tangentially propagating jet shower low $p_{T}$ partons from the jet shower and
Mach-cone-like excitation are clearly deflected by both the density gradience and the radial flow, giving rise to 
the azimuthal distributions that peak at an angle away from the initial jet direction.  For jet showers 
that travel against the radial flow, the same deflection splits the azimuthal
distribution of low $p_{T}$ partons to become a double-peaked one. 

\begin{figure}
\centerline{\includegraphics[width=7.5cm]{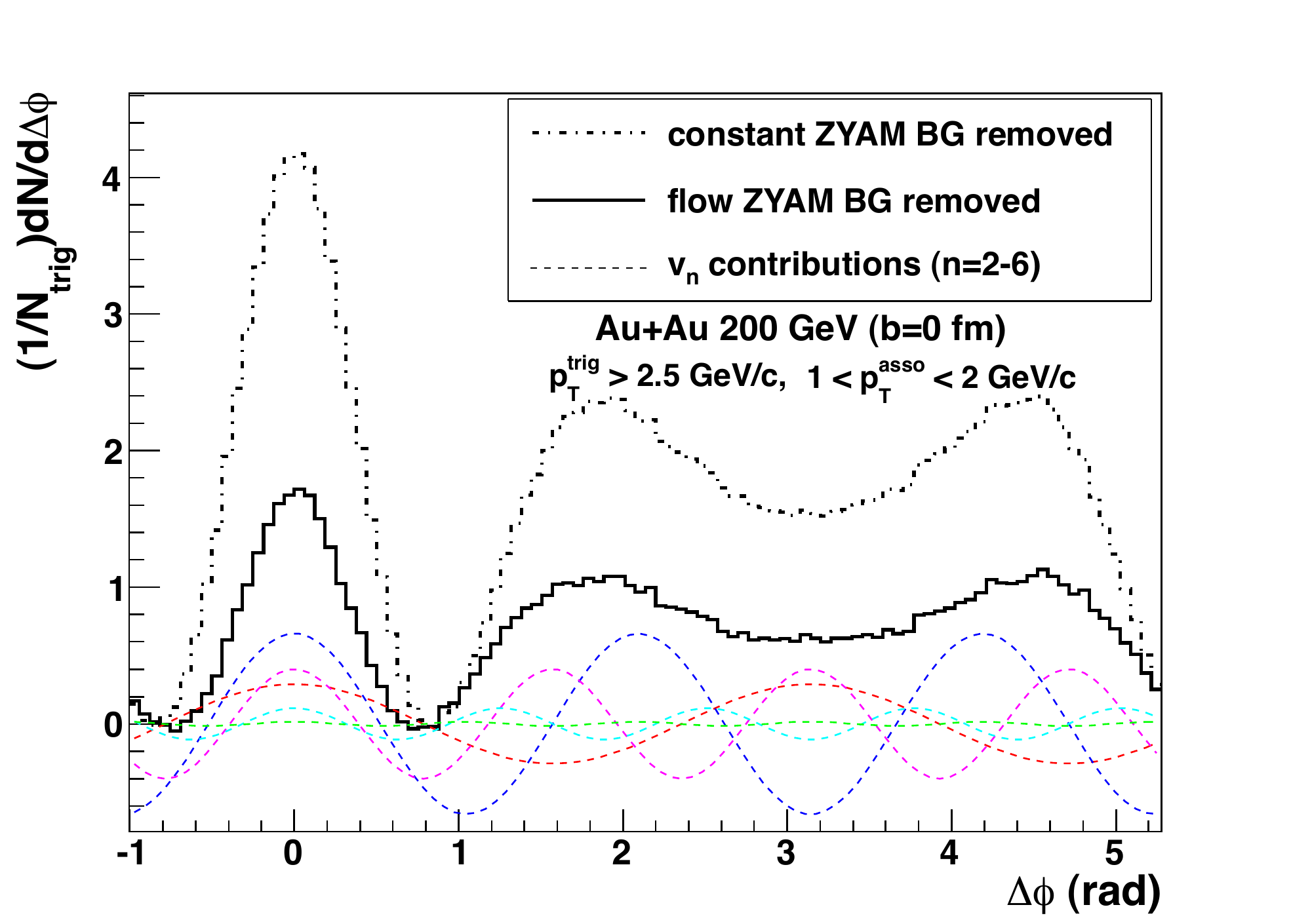}\includegraphics[width=7.5cm]{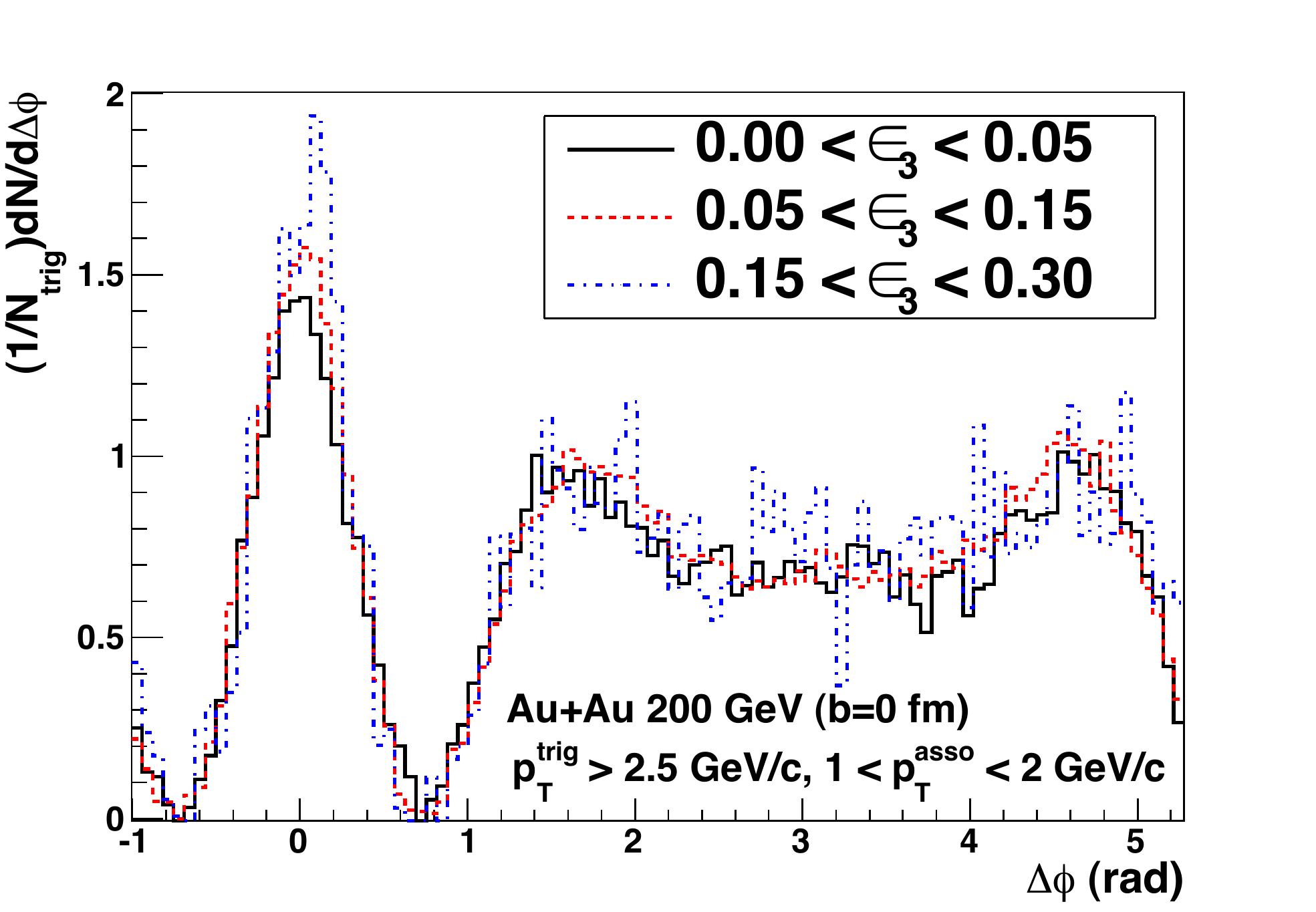}}
\caption{(Color online) (left) AMPT results on dihadron azimuthal correlation before (dot-dashed) and after (solid)
subtraction of contribution from harmonic flow $v_{n} (n=2-6)$ for $p_{T}^{\rm trig}>2.5$ GeV/$c$ 
and $1 < p_{T}^{\rm asso} < 2$ GeV/$c$. (right) Dihadron correlations after subtraction of harmonic flow
with different values of geometric triangularity $\epsilon_{3}$ for $p_{T}^{\rm trig}>2.5$ GeV/$c$ 
and $1 < p_{T}^{\rm asso} < 2$ GeV/$c$ .}
 \label{fig-dih1}
\end{figure}

Multiple scatterings in heavy-ion collisions lead to fluctuation in local parton number density or hot spots 
from both soft and hard interactions. Shown in the right panel of Fig.~\ref{fig1} is a contour plot of initial parton 
density distribution in transverse plane $dN/dxdy$ of a typical AMPT event for central ($b=0$) $Au+Au$
collisions at $\sqrt{s}=200$ GeV.  The irregular distribution of hot spots will lead to harmonic flow due to
collection expansion. The contributions from these harmonic flows to dihadron correlations can be calculated as
\begin{equation}
f(\Delta \phi ) = B\left(1 + \sum\limits_{n = 1}^\infty {2\langle v_n^{\rm trig} v_n^{\rm asso}\rangle \cos n\Delta \phi} \right),  \label{eq:BG}
\end{equation}
where B is a normalization factor determined by the ZYAM (zero yield at minimum) scheme of
background subtraction , $v_n^{trig}$ and $v_n^{asso}$ 
are harmonic flow coefficients for trigger and associated hadrons.  
For the study of jet-induced medium
excitation, it is important to isolate and subtract contributions from harmonic flows, especially the triangular flow,
since it contributes the most to the double-peak structure of back-to-back dihadron correlation. 
Shown in the right panel of Fig.~\ref{fig-dih1}
are dihadron correlations before (dot-dashed) and after (solid) the removal of
contributions from harmonic flows for $p_{T}^{\rm trig}>2.5$ GeV/$c$ 
and $1 < p_{T}^{\rm asso} < 2$ GeV/$c$. Also shown in the figure are contributions from each harmonic flow $n=2$-6 (dashed). 
These contributions are significant for up to $n=5$ harmonics.

As seen in the left panel of Fig.~\ref{fig-dih1}, dihadron correlation after subtraction of contributions from harmonic flows still has
a double-peak feature on the away-side of the trigger due to  jet-induced medium excitation and hadrons 
from expanding hot spots. The structure  should
be intrinsic to the jet-induced medium excitation and hot spots themselves and insensitive to the fluctuation of the initial geometry of the dense matter at a fixed
impact-parameter. As shown in the right panel of Fig.~\ref{fig-dih1}, the dihadron correlations after subtraction of
contributions from harmonic flows become independent on the initial geometric triangularity $\epsilon_{3}$.

\begin{figure}
\centerline{\includegraphics[width=7.5cm]{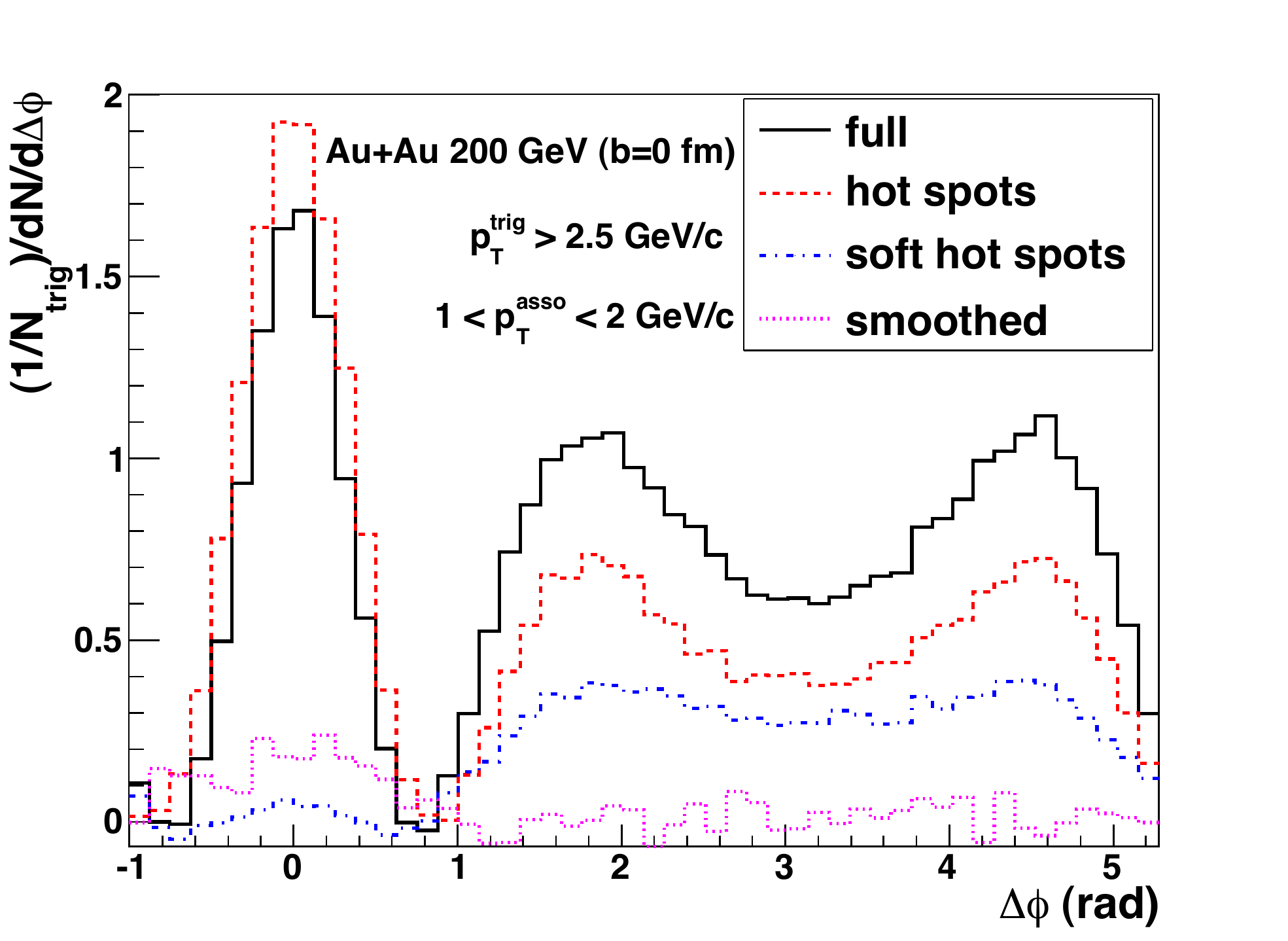}\includegraphics[width=7.95cm]{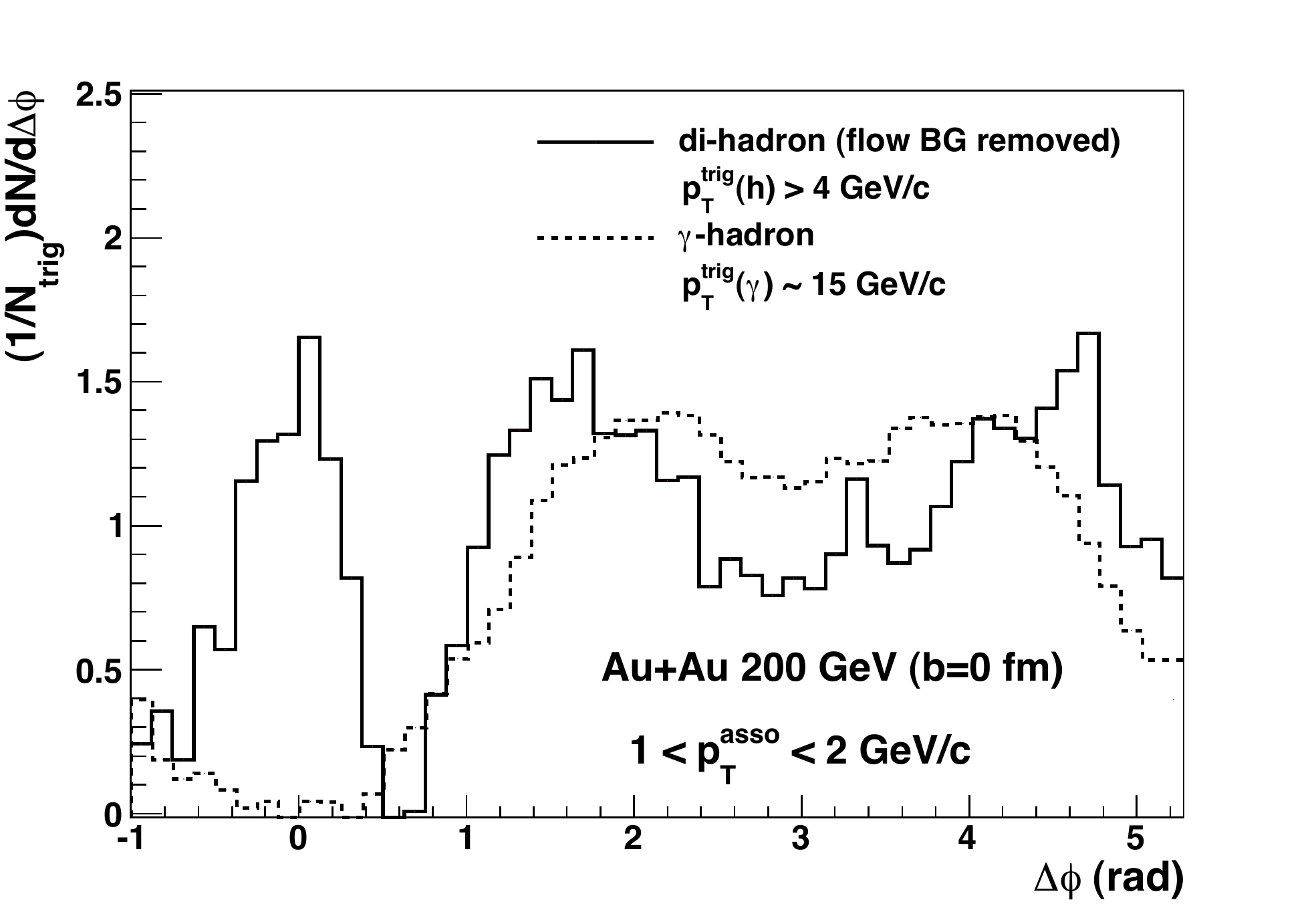}}
\caption{(Color online) (left)  Dihadron correlation (with harmonic flow subtracted) from AMPT with different
initial conditions for $p_{T}^{\rm trig}>2.5$ GeV/$c$ and $1 < p_{T}^{\rm asso} < 2$ GeV/$c$. 
See text for details on the different initial conditions. (right) Dihadron correlation (solid)
compared with $\gamma$-hadron correlation (dashed) from AMPT for $p_{T}^{\rm trig}(h)>4$ GeV/$c$,
$p_{T}^{\rm trig}(\gamma)\ge 15$ GeV/$c$ and $1 < p_{T}^{\rm asso} < 2$ GeV/$c$ .}
 \label{fig-dih3}
\end{figure}

To study the structure of dihadron azimuthal correlation from jets and hot spots separately, we successively 
switch off each mechanism in AMPT model calculations. By randomizing the azimuthal angle of each jet shower
parton in the initial condition from HIJING simulations, we effectively switch off the initial
back-to-back correlation of dijets. The dihadron correlation (dashed) after subtraction of hadronic flows denoted as ``hot spots'' in the
left panel of Fig.~\ref{fig-dih3} still exhibits a double-peak on the away-side that comes only from hot spots. 
It has roughly the same opening angle $\Delta\phi\sim 1$ (rad) as in the full simulation (solid).  However, the magnitude
of the double-peaked away-side correlation is reduced by about 40\%, which can be attributed to
dihadron from medium modified dijets and jet-induced medium excitation. 
When we turn off jet production in the HIJING initial condition, soft partons from string materialization
can still form ``soft hot spots''  that lead to a back-to-back dihadron correlation (dot-dashed) with a weak
double-peak. Jet shower partons increase the parton density in ``hot spots'' and lead
to a stronger double-peak dihadron correlation than that of ``soft hot spots''.
Without jets in AMPT, one can further randomize the polar angle of  transverse coordinates of
soft partons and therefore eliminate the ``soft hot spots''. The dihadron correlation from such smoothed initial
condition becomes almost flat (dotted).

Since high-$p_{T}$ $\gamma$'s do not interact with the dense medium, their emission should be uniform
in the azimuthal angle and uncorrelated with the harmonic flow and collective flow of the hot spots. Therefore,
$\gamma$-hadron correlation should only come from $\gamma$-triggered jets and their shape should
reflect directly the medium modification of the jets and jet-induced medium excitation.
Shown in the right panel of Fig.~\ref{fig-dih3} are dihadron correlations
(solid) after subtraction of harmonic flow as compared with $\gamma$-hadron
correlation. The two correlations are comparable in magnitude but dihadron has a more pronounced 
double-peak which can be attributed to the addition of dihadrons from hot spots and the
geometric bias toward surface and tangential emission that enhances deflection
of jet showers and jet-induced medium excitation \cite{Li:2010ts} by the radial flow.
Such difference is important to measure in experiments
that will provide critical information on jet-induced medium excitation and evolution of hot
spots in high-energy heavy-ion collisions.

This work is supported
by the NSFC of China under Projects Nos. 10610285, 10635020, 10705044, 10825523, 10975059,
11035009, the Knowledge Innovation Project of Chinese Academy of Sciences under Grant No. KJCX2-EW-N01
and  by the U.S. DOE under Contract No. DE-AC02-05CH11231 and within the framework of the JET Collaboration.





\bibliographystyle{elsarticle-num}



\end{document}